\documentclass[sigconf]{acmart}

\usepackage{algorithm}
\usepackage{algorithmic}
\usepackage{array}
\usepackage{amsfonts}
\usepackage{amssymb} 
\usepackage{amsmath}
\usepackage{bm}
\usepackage{booktabs}
\usepackage{color}
\usepackage{epstopdf}
\usepackage{enumitem}
\usepackage{footmisc}
\usepackage{graphicx}
\usepackage{hyperref}
\usepackage{fancyhdr}
\usepackage{listings}
\usepackage{multirow}
\usepackage{mathrsfs}
\usepackage{pifont}
\usepackage{subfigure}
\usepackage{setspace}
\usepackage{soul}
\usepackage{textcomp}
\usepackage{url}
\usepackage{xcolor}
\usepackage{xspace}
\usepackage[normalem]{ulem}
\usepackage[np, autolanguage]{numprint}
\usepackage[skip=0pt]{caption}
\newcommand{\paratitle}[1]{\vspace{1.5ex}\noindent\textbf{#1}}
\newcommand{\ie}{\emph{i.e.,}\xspace}

\newcommand{\eg}{\emph{e.g.,}\xspace}

\newcommand{\ignore}[1]{}

\newcommand{\eat}[1]{}

\ignore{\makeatletter
\g@addto@macro\normalsize{%
  \abovedisplayskip 3pt plus 2pt minus 2pt%
  \belowdisplayskip \abovedisplayskip
  \abovedisplayshortskip 3pt plus2pt  minus2pt%
  \belowdisplayshortskip 3pt plus2pt minus2pt%
}

\makeatother
}

\copyrightyear{2020}
\acmYear{2020}
\setcopyright{acmcopyright}\acmConference[SIGIR '20]{Proceedings of the 43rd International ACM SIGIR Conference on Research and Development in Information Retrieval}{July 25--30, 2020}{Virtual Event, China}
\acmBooktitle{Proceedings of the 43rd International ACM SIGIR Conference on Research and Development in Information Retrieval (SIGIR '20), July 25--30, 2020, Virtual Event, China}
\acmPrice{15.00}
\acmDOI{10.1145/3397271.3401111}
\acmISBN{978-1-4503-8016-4/20/07}

\begin{document}

\title{Sequential Recommendation with Self-Attentive Multi-Adversarial Network}

\author{Ruiyang Ren$^{1,4}$, Zhaoyang Liu$^2$, Yaliang Li$^2$, Wayne Xin Zhao$^{3,4*}$, Hui Wang$^{1,4}$,}
\thanks{$^*$Corresponding author.}
\author{Bolin Ding$^2$, and Ji-Rong Wen$^{3,4}$}
\affiliation{
 \institution{$^1$School of Information, Renmin University of China}
 \institution{$^2$Alibaba Group}
 \institution{$^3$Gaoling School of Artificial Intelligence, Renmin University of China}
 \institution{$^4$Beijing Key Laboratory of Big Data Management and Analysis Methods}
}
\affiliation{%
  \institution{\{reyon.ren, hui.wang, jrwen\}@ruc.edu.cn, \{jingmu.lzy, yaliang.li, bolin.ding\}@alibaba-inc.com, batmanfly@gmail.com}
}

\renewcommand{\shortauthors}{Ruiyang Ren, Zhaoyang Liu, Yaliang Li, Wayne Xin Zhao, Hui Wang, Bolin Ding and Ji-Rong Wen}

\begin{abstract}
Recently, deep learning has made significant progress  in the task of sequential recommendation.
Existing neural sequential recommenders typically adopt a generative way  trained with Maximum Likelihood Estimation~(MLE).
When context information (called \emph{factor}) is involved, it is difficult to analyze when and how each individual factor would affect the final recommendation performance.



For this purpose, we take a new perspective and introduce adversarial learning to sequential recommendation. In this paper, 
we present a Multi-Factor Generative Adversarial Network~(MFGAN) for explicitly modeling the effect of context information on sequential recommendation.
 Specifically, our proposed MFGAN has two kinds of modules: a Transformer-based generator taking user behavior sequences as input to recommend the possible next items, and multiple factor-specific discriminators to evaluate the generated sub-sequence from the perspectives of different factors. To learn the parameters, we adopt the classic policy gradient method, and utilize the reward signal of discriminators for guiding the learning of the generator.
Our framework is flexible to incorporate multiple kinds of factor information, and is able to trace how each factor contributes to the recommendation decision over time.  
Extensive experiments conducted on three real-world datasets demonstrate the superiority of our proposed model over the state-of-the-art methods, in terms of effectiveness and interpretability.
\end{abstract}



\maketitle

\section{Introduction}
\label{sec:intro}

Recommender systems aim to accurately characterize user interests and provide personalized recommendations in a variety of real-world applications. They serve as an important information filtering technique  to alleviate the information overload problem and enhance user experiences. 
In most applications, users' interests are dynamic and evolving over time.
It is essential to  capture the dynamics of sequential user behaviors for making appropriate recommendations.

In the literature, various methods \cite{RendleFS10, GRU4Rec, kang2018self} have been proposed for sequential recommender systems. 
Early methods usually utilize the Markov assumption that the current behavior is tightly related to the previous ones~\cite{RendleFS10}.
Recently, sequential neural networks such as recurrent neural network~\cite{GRU} and Transformer~\cite{transformer} have been applied to recommendation tasks as these networks can  characterize sequential user-item interactions and  learn effective representations of user behaviors~\cite{GRU4Rec, kang2018self}. Besides, several studies have proposed to incorporate context information to enhance the  performance of neural sequential recommenders~\cite{GRU-f, li2019review, HuangWSDM19}. The advantages of these sequential neural networks have been experimentally confirmed as they have achieved significant performance improvements.

Typically,  existing neural sequential  recommenders~\cite{kang2018self, GRU4Rec} adopt a generative way to predict future items and learn the parameters using Maximum Likelihood Estimation~(MLE).
However, it has been found that MLE-based training is easy to suffer from issues such as data sparsity or exposure bias~\cite{Ranzato2015Sequence,Yu2017SeqGAN} in sequence prediction.
Especially, in such an approach,  when context information  (called as \emph{factor} in this paper) is incorporated, it has to be integrated with the original sequential prediction component~\cite{GRU-f,HuangWSDM19, li2019review}. 
The consequence is that  various factors (\eg price and brand of a product in the e-commerce scenario) are either mixed in the sequential context representations, or coupled with the black-box recommendation module.  Therefore, we cannot accurately figure out when and how each individual factor would affect the final recommendation performance. 
These disadvantages weaken or even impede their applications in a wide range of  decision making scenarios. 
It is important to explicitly and effectively characterize the effect of various factors in sequential recommender systems. 

\ignore{More importantly, when context information  (called as \emph{factor} in this paper) is utilized, it has to be integrated with the original sequential prediction component~\cite{GRU-f,HuangWSDM19, li2019review}.
In such  a way,  various factors (\eg price and brand of a product in the e-commerce scenario) are either mixed in the sequential context representations, or coupled with the black-box recommendation module.  
Therefore, we cannot accurately figure out when and how each individual factor would affect the final recommendation performance. 
These disadvantages weaken or even impede their applications in a wide range of business decision making. 
It is essential to be able to explicitly and effectively characterize the effect of various factors on users' sequential behaviors. 
}

\ignore{
We argue that it is not the most suitable to seek interpretable recommendation with available factor information.    
However, these sequential neural networks also pose great challenges in understanding the sequential characteristics of user behaviors due to the black-box nature of deep neural networks.
Even with supplementary context information (called as \emph{factor} in this paper), it is still difficult to understand the reasons behind the recommendation outputs.
In existing context-aware sequential recommendation algorithms~\cite{GRU-f, HuangWSDM19, li2019review, chen2016learning},
various factors, such as price and brand of a product in the e-commerce scenario, or genres and actors of a movie in the movie scenario, are either mixed in the sequential context representations, or coupled with the black-box recommendation module.  
}



In the light of this challenge, we propose to use an adversarial training approach to developing sequential recommender systems. 
Indeed, the potential advantage of Generative Adversarial Network~(GAN) has been shown in  collaborative filtering methods~\cite{CFGAN,IRGAN}. Different from prior studies, our novelty is to  decouple  \emph{factor utilization} from the \emph{sequence prediction} component via adversarial training.
Following the GAN framework~\cite{GAN-2014-NIPS}, we set two different components, namely generator and discriminator. 
In our framework, 
the generator predicts the future items for recommendation relying  on user-item interaction data alone, while the discriminator  judges the rationality of the generated recommendation sequence based on available  information of various factors.
Such an approach allows more flexibility in utilizing external context information in sequential recommendation, which is able to improve the recommendation interpretability. 

To this end, in this paper, we present  a novel \emph{Multi-Factor Generative Adversarial Network (MFGAN)}. 
Specifically, our proposed MFGAN has two essential kinds of modules: (1) a Transformer-based generator taking user behavior sequences as input to recommend the possible next items, and (2) multiple factor-specific discriminators to evaluate the generated recommendations from the perspectives of different factors.
Unlike the generator,  the discriminator adopts a bi-directional Transformer-based architecture, and it can refer to the information of subsequent positions for sequence evaluation.
In this way, the discriminator is expected to make more reliable judgement by considering the overall sequential characteristics  \emph{w.r.t.} different factors.
Due to the discrete nature of item generation, 
the training of the proposed MFGAN method is realized in a reinforcement learning way by policy gradient. 
The key point is that we utilize the discriminator modules to provide the reward signal for guiding the learning of the generator.

Under our framework, various factors are decoupled from the generator, and they are utilized by the discriminators to derive supervision signals to improve  the generator. 
To validate the effectiveness of the proposed MFGAN, we conduct extensive experiments on three real-world datasets from  different domains. Experimental results show that the proposed MFGAN is able to achieve better performance compared to several competitive methods. 
We further show  the multi-adversarial architecture  is indeed useful to stabilize the learning process of our approach.
Finally, qualitative analysis demonstrates that the proposed MFGAN can explicitly characterize the effect of various factors over time for sequential recommendation, making the recommendation results highly interpretable.

\eat{
Our main contributions are summarized as follows:
\begin{itemize}[leftmargin=0.1cm,labelsep=.5em, noitemsep]
    \item To the best of our knowledge, we are the first to introduce adversarial training into the sequential recommendation task, unifying the unidirectional generative process and bidirectional discrimination for the recommendation.
    \item We propose a multi-discriminator structure that can decouple different factors and improve the performance of sequential recommendation. We  analyze the effectiveness and the stability of the multi-adversarial architecture in our task.
    \item Extensive experiments conducted on three real-world datasets demonstrate the benefits of the proposed MFGAN over state-of-the-art methods.
\end{itemize}
}

Our main contributions are summarized as follows:

$\bullet$ To the best of our knowledge, we are the first to introduce adversarial training into the sequential recommendation task, and design the unidirectional generator for prediction and bidirectional discriminator for evaluation.

$\bullet$ We propose a multi-discriminator structure that can decouple different factors and improve the performance of sequential recommendation. We  analyze the effectiveness and the stability of the multi-adversarial architecture in our task.

$\bullet$ Extensive experiments conducted on three real-world datasets demonstrate the benefits of the proposed MFGAN over state-of-the-art methods, in terms of both effectiveness and interpretability.


\section{Problem Definition}
\label{sec:definition}

In this section, we first formulate the studied problem of sequential recommendation before diving into the details of the proposed method.
Let $\mathcal{U}$ and $\mathcal{I}$ denote a set of users and items, respectively, where $|\mathcal{U}|$ and $|\mathcal{I}|$ are the numbers of users or items.
Typically, a user $u$ has a chronologically-ordered interaction sequence of items: $\{i_1, i_2, \dots, i_t, \dots, i_{n}\}$, where $n$ is the total number of interactions and $i_t$ is the $t$-th item that user $u$ has interacted with. 
For convenience, we use $i_{j:k}$ to denote the subsequence of the entire sequence, \ie  $i_{j:k}=\{i_j,  \dots, i_k\}$, where $1 \le j < k \le n $.
 Besides, we assume that each item $i$ is associated with $m$ kinds of contextual information, corresponding 
to $m$ factors, \eg artist, album and popularity in music recommender system. 
 
\ignore{
The historical behaviors of each user is denoted by $S_u=\{i_1, i_2, \dots, i_t, \dots, i_{n}\}$ where $i_t \in \mathcal{I}, t \in [1,|u|]$ and $|u|$ is the total number of the user behaviors until the timestamp $t$.
Item $i$ also has a set of features such as product category, movie genre, and music popularity. These features are the factors we discussed above. 
All these features ($i.e.$, factors) are only available for the discriminators in the proposed method as we want to decouple mixed factors through the designed multi-adversarial architecture. More details will be discussed in the next section.
}

Based on the above notations, we now define the task of sequential recommendation.
Formally, given the historical behaviors of a user (\ie $\{i_1, i_2, \dots, i_t, \dots, i_{n}\}$) and the context information of items, our task aims to predict the next item that she/he is likely to interact with at the $(n+1)$-th step.

\section{Methodology}
\label{sec:method}

In this section, we first give an overview of the proposed \emph{Multi-Factor Generative Adversarial Network}~(MFGAN) framework, and then introduce the design of the generator and discriminators.
The details of the training process are also discussed in this section.

\begin{figure*}[ht]
	\centering 
	\includegraphics[width=0.76\textwidth]{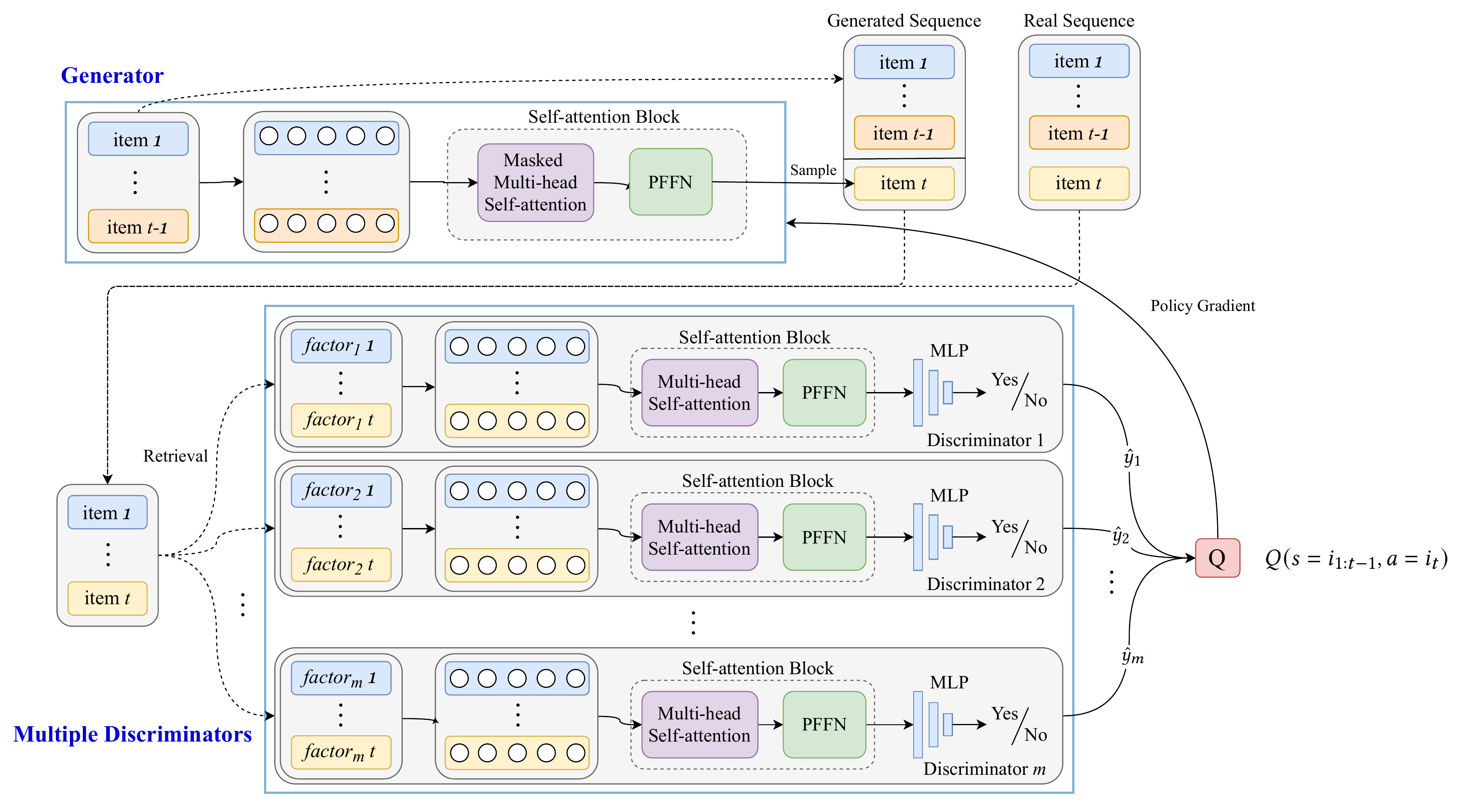}
	\caption{The overview of the proposed MFGAN model consisting of a generator and multiple discriminators. The upper  and the bottom framed parts correspond to the generator and multi-discriminator components, respectively. } 
	\label{fig-model} 
\end{figure*}

\subsection{Multi-Factor Generative Adversarial Network Framework}

Figure~\ref{fig-model} presents the overview of our proposed MFGAN framework for sequential recommendation.

\subsubsection{Basic Components} In this framework, we have two kinds of components undertaking different responsibilities for sequential recommendation:

(1) The upper component is the prediction component ($i.e.$ generator $G$) which is a sequential recommendation model and successively generates the next items based on the current historical sequence.
Note that the generator will not use any context information from the item side. It only makes the prediction conditioned on historical sequence data.

(2) The lower component is the evaluation component that is a set of $m$ discriminators $\{D_1, D_2, \dots, D_m\}$ for judging the rationality of generated sequences by using the information from multiple perspectives.
Each discriminator performs the judgement from a certain perspective based on the  information of some corresponding factor.
For example, in music recommender system, we may have multiple discriminators  specially designed with category information, popularity statistics, artist and album of music, respectively.

\subsubsection{Overall Procedure}  Following standard GAN~\cite{GAN-2014-NIPS},  the generator and multiple discriminators will play a min-max game.
At the $t$-step, the generator first generates a predicted item $\hat{i}_t$ based on the historical  sequence $\{ i_1, \dots, i_{t-1} \}$.
Then, 
each discriminator takes the $t$-length  sequence $\{ i_1, \dots, i_{t-1}, \hat{i}_t\}$ as the input and evaluates the rationality of the generated sequence using the information of some factor. The evaluation results are sent back to the generator to guide the learning of the generator at the next round.
Correspondingly, the discriminator is updated by taking the generated sequence and actual sequence~(\ie ground-truth user behaviors) as the training samples for improving its discriminative capacity.
As such, the two components force each other to improve in a mutual reinforcement way.

\subsubsection{Merits} There are three major merits of using such a framework for sequential recommendation.
First, generally speaking, it is relatively difficult to train a capable generation-based sequential recommender using a direct optimization with a maximum likelihood loss (\eg exposure bias or data sparsity~\cite{Ranzato2015Sequence}). We utilize the discriminators to monitor the quality of the recommendation results of the generator, which are able to gradually improve the final recommendation performance. 
Second, it is more flexible to incorporate various kinds of factor information into  discriminators, so that the generator can  focus on the generation task itself.
Such a way is more resistible  to useless or noisy information from context data. It is easier to incorporate additional factors into an existing model. 
 Third, instead of modeling all the factors in a discriminator, our framework decouples the effect of each factor by using multiple discriminators, which also improves the interpretability (\eg explaining why a special behavioral transition occurs) of the generated sequences.

To instantiate the framework, we adopt the recently proposed self-attentive neural architecture~(\eg Transformer~\cite{transformer}) to develop the components of generator and discriminators, since it has been shown to be successful in various sequence-oriented tasks, including sequential recommendation~\cite{kang2018self}. While, it is flexible to implement our framework with other classes of models in practice.    
In the following sections, we will introduce the details of both components.

\subsection{The Generator Component}
\label{sec-generator}
In the MFGAN framework, let $G_\theta$ denote the generator component, parameterized by $\theta$, where $\theta$ denotes the set of all the related parameters in $G$.
We develop the  generator for sequential recommendation model by stacking the embedding layer, self-attention block, and the prediction layer to generate the target items. Next, we describe each part in detail.

\subsubsection{Embedding Layer} 
\label{embedding-layer}
We maintain an item embedding matrix $\bm{M}_G \in \mathbb{R}^{|\mathcal{I}| \times d}$ to project original one-hot representations of items to $d$-dimensional dense representations.
Given a $n$-length sequence of historical interactions, we apply a look-up operation from $\bm{M}_G$ to form the input embedding matrix $\bm{E} \in \mathbb{R}^{n \times d}$.
Furthermore, we incorporate a learnable position encoding matrix $\bm{P} \in \mathbb{R}^{n \times d}$ to enhance the input representations. 
In this way,  the input representations $\bm{E}_G \in \mathbb{R}^{n \times d}$ for the generator can be obtained by summing two embedding matrices:  $\bm{E}_G=\bm{E}+\bm{P}$.

\ignore{As for input user behaviors, we maintain an embedding matrix $M_G \in \mathbb{R}^{|\mathcal{I}| \times d}$ with the dimensionality of $d$ to project the high-dimensional one-hot representation of the item to low-dimensional dense representations.
After stacking the input by the embedded representation from $M_G$, we get an input embedding matrix $E \in \mathbb{R}^{n \times d}$.
A learnable position encoding $P$ is also added to $E$ and supply the temporal information for the self-attention mechanism. Finally, the input of the generator is denoted by:
\begin{equation}
\begin{split}
    \label{eq-embedding}
    E_G = 
    \begin{bmatrix}
    E_1 + P_1\\
    E_2 + P_2 \\
    \vdots \\
    E_n + P_n
    \end{bmatrix}.
\end{split}
\end{equation}
}

\subsubsection{Self-attention Block} Based on the embedding layer, we stack multiple self-attention blocks. 
A self-attention block generally consists of two sub-layers, a multi-head self-attention sub-layer and a point-wise feed-forward network.
Instead of attending information of user sequences with a single attention function,
multi-head self-attention mechanism has been adopted for effectively extracting the information  from different representation subspaces.
Specifically, multi-head self-attention is defined as below:

\begin{equation}
\label{eq-attention}
\begin{split}
\text{MultiHeadAtt}(\bm{F}^l) &= [head_i;head_2;\dots,head_h]\bm{W}^O, \\
head_i &= \text{Attention}(\bm{F}^l \bm{W}^Q_i, \bm{F}^l \bm{W}^K_i, \bm{F}^l \bm{W}^V_i),
\end{split}
\end{equation}
where the $\bm{F}^l$ is the input for the $l$-th layer.
When $l=0$, we set $\bm{F}^l$ to the input $\bm{E}_G$ of the generator.
Let $h$ denote the number of heads. 
The projection matrix $\bm{W}^Q_i \in \mathbb{R}^{d \times d/h}, \bm{W}^K_i \in \mathbb{R}^{d \times d/h}, \bm{W}^V_i \in \mathbb{R}^{d \times d/h}$ and $\bm{W}^O \in \mathbb{R}^{d \times d}$ are the corresponding learnable parameters for each attention head.
The attention function is implemented by scaled dot-product operation:

\begin{equation}
    \text{Attention}(\mathbf{Q}, \mathbf{K}, \mathbf{V}) = \text{softmax}(\frac{\mathbf{Q} \mathbf{K}^T}{\sqrt{d/h}})\mathbf{V},
\end{equation}
where $\mathbf{Q}=\bm{F}^l \bm{W}^Q_i,\mathbf{K}=\bm{F}^l \bm{W}^K_i$, and $\mathbf{V}=\bm{F}_l \bm{W}^V_i$, which are the linear transformations of the input embedding matrix.
The temperature $\sqrt{d / h}$ is the scale factor to avoid large values of the inner product.
In sequential recommendation, we can only utilize the information before  current time step, and we apply  the mask operation to the output of the multi-head self-attention function, removing  all connections between $\mathbf{Q}_i$ and $\mathbf{K}_j$ (for all cases of $j > i$).

As shown in Eq.~\eqref{eq-attention}, the multi-head attention function is mainly built on the linear projections.
We further endow the non-linearity of the self-attention block by applying a point-wise feed-forward network as:

\begin{align}
    \text{PFFN}(\bm{F}^l) &= [\text{FFN}(\bm{F}^l_1)^\top; \cdots; \text{FFN}(\bm{F}_n^l)^\top],\\
    \text{FFN}(\bm{x}) &= \max(0, \bm{x}\bm{W}_1 + \bm{b}_1) \bm{W}_2 + b_2,
\end{align}
where $\bm{W}_1, \bm{b}_1, \bm{W}_2, b_2$ are the trainable parameters and not shared across layers.

\eat{
The stacking of the transformer block is also applied.
As the model goes deep, it's usually beneficial to model more and more complex item-item interactions.
To alleviate the overfitting problem, we add residual connection~\cite{he2016deep} between each sub two layers intra the transformer structure, followed by layer normalization~\cite{layer_norm} and dropout operation~\cite{dropout}.
}

\subsubsection{Prediction Layer}

At the final layer  of the generator, we calculate the user's preference over the item set  through the softmax function:
\begin{equation}
\begin{split}
    \label{eq-predict}
    G_{\theta}(i_t | i_{1:t-1}) = \text{softmax}(\bm{F}_n^L \bm{M_G}^\top)_{[i_t]},
\end{split}
\end{equation}
where $L$ is the number of self-attention blocks and $\bm{M}_G$ is the maintained item embedding matrix defined in Section~\ref{embedding-layer}. 



\ignore{
In the final layer  of the generator, we calculate the user’s preference for items through a dot product operation.
\begin{equation}
\begin{split}
    \label{eq-predict}
    \bm{y}_{G} = \bm{F}^L \bm{M_G}^\top,
\end{split}
\end{equation}
where $ \bm{y}_{G} \in \mathbb{R}^{|\mathcal{I}|}$ is the preference vector over the item set, and  $L$ is the number of transformer blocks and $\bm{M}_G$ is the maintained item embedding matrix defined in Section 3.2.1.
The item with the highest prediction value in $\bm{y}_{G} $ is selected as the generated item for the generator.
}

\subsection{Factor-specific Discriminator Components}
As mentioned before, we consider $m$ kinds of factor information that is useful to improve the sequential recommendation. Instead of directly feeding them into the generator, we set a unique discriminator for each factor, such that various kinds of context information  can be utilized and decoupled via the factor-specific discriminators. 

Specially, we have $m$ discriminators $D_\Phi = \{ D_{\phi_1}, D_{\phi_2}, \dots, D_{\phi_m} \}$, in which the $j$-th discriminator is parameterized by $\phi_j$.
The function of each discriminator is to determine whether the generated recommendation sequence by the generator is rational or not. This is cast as a binary classification task, \ie discriminating between generated or actual recommendation sequence. 
We assume that different discriminators are equipped with different parameters and work independently. 

\ignore{As for discriminator, we propose a multiple $\Phi$-parameterized discriminative model $D_\Phi = \{ D_{\phi_1}, D_{\phi_2}, \dots, D_{\phi_m} \}$ to provide a multi-perspective guidance for improving generator $G_\theta$. 
The purpose of a single discriminator is to determine if a given sequence is reasonable or not from the corresponding factor information.
Different discriminators work independently and do not share parameters.
}

\ignore{Specifically, each discriminator is designed for a binary classification task built on self-attention blocks. 
Since the task of the discriminator is much easier than that of the generator, we have a simplified architecture of the discriminator with only one self-attention block to prevent the discriminator from being too critical and further affecting the training process of the generator.
}

\subsubsection{Embedding Layer} Considering a specific discriminator $D_{\phi_j}$,  we first construct an  input embedding matrix $\bm{E}_{D}^{j}\in \mathbb{R}^{n\times d}$ for a $n$-length sequence by summing the factor-specific embedding matrix $\bm{C}^{j}$ and the positional encoding matrix $\bm{P}$, namely $\bm{E}_{D}^{j} = \bm{C}^{j}+\bm{P}$. 
To construct the  $\bm{C}^{j}$, we adopt a simple yet effective method: first discretize the possible values of a factor into several bins,  then set a unique embedding vector for each bin, and finally derive $\bm{C}^{j}$ using a look-up operation by concatenating the embeddings for the bin IDs from the input sequence.  

\subsubsection{Architecture}  To develop the discriminator, we adopt the similar architecture of the generator. In our framework, the generator predicts the recommended sequence, and the discriminators are mainly used to improve the generator. 
Hence, we adopt a relatively weak architecture with only one self-attention block for avoiding the case that the discriminator is too strong and cannot send suitable feedback to the generator. 
The one-layer self-attention block is computed as:

\begin{align}
    \label{eq-sa-d}
    \bm{A}^{j} &= \text{MultiHeadAtt}(\bm{E}_{D}^{j}), \\
    \bm{H}^{j} &= \text{PFFN}(\bm{A}^{j}).
\end{align}
Note that unlike the generator, the self-attention block of the discriminator can refer to the information of subsequent positions when trained at the $t$-th position.
Hence,  the discriminator adopts a \emph{bi-directional} architecture by removing the mask operation. In this way, the discriminator can model the interaction between any two positions, and make a more accurate judgement by considering the overall sequential characteristics. 
While, the generator does not utilize such bi-directional sequential characteristics. 
As such, 
the discriminator is expected to provide additional supervision signals, though it shares the similar architecture with the generator. 


\ignore{Note that unlike the generator, the self-attention block in discriminator can consider the subsequent positions when training the $t$-$th$ position, because the task of the discriminator is no longer to predict the current position based on the previous positions, but to judge the rationality throughout the sequence.
We remove the mask operation for each discriminator and keep the interactions between any two positions.}

Finally, the  degree of the rationality  for the generated recommendation sequence is measured by a  Multiple-Layer Perceptron~(MLP):

\begin{equation}\label{eq-mlp}
    \hat{y}_{j} = \text{MLP}(\bm{H}_n^j),
\end{equation}
where $\hat{y}_{j}$ is the predicted degree of the rationality from the the  MLP component based on 
 the output of the self-attention block $\bm{H}^j$.
A  rationality score reflects the probability that a sequence is from actual data distribution judged by some discriminator. 

Since we have $m$ discriminators  \emph{w.r.t.} different factors, we can obtain a set of predicted rationality scores $\{ \hat{y}_{1}, \hat{y}_{2}, \dots, \hat{y}_{m} \}$. 
As will be illustrated later, these rationality scores can be used for supervision signals to guide the learning of the generator.

\subsection{Multi-adversarial Training}
As described previously, there is  one generator $G_\theta$ and multiple discriminators $D_\Phi = \{ D_{\phi_1}, D_{\phi_2}, \dots, D_{\phi_n} \}$. The generator $G_\theta$ successively predicts the next item based on historical sequence data, and the discriminators try to discriminate between the predicted sequence and the actual sequence.  In this part, we present the multi-adversarial training algorithm for our approach. 

\ignore{
As illustrated in previous subsection, we have one generative model 
$G_\theta$ and multiple discriminator $D_\Phi = \{ D_{\phi_1}, D_{\phi_2}, \dots, D_{\phi_n} \}$.
Generator $G_\theta$ successively generates the next item based on user historical sequence $I = \{ i_1, \dots, i_{T-1}\}$ and correspondingly yield the following items $I = \{ i_2, \dots, i_T\}$ at each time step.
Multi-discriminator $D_\Phi$ can determine whether the sequence generated by the generator is a reasonable sequence or not from multiple perspectives with different factors.
}

\subsubsection{RL-based Formalization} Because sampling from the item set is a discrete process,  gradient descent cannot be directly applied to solve the original GAN formulation for our recommendation task. 
As such, following~\cite{Yu2017SeqGAN},  we first formalize the sequential recommendation task in a reinforcement learning~(RL) setting. At the $t$-step, the \emph{state} $s$ is represented by the previously recommended sub-sequence  $i_{1:t-1}=\{ i_1, i_2, \dots, i_{t-1} \}$; the \emph{action} $a$ is to select the next item $i_t$ for recommendation, controlled by a \emph{policy} $\pi$ that is defined according to the generator: $\pi(a=i_t | s)=G_\theta(i_t| i_{1:t-1})$; when an action is taken, it will transit from $s_t$  to a new state $s'$, corresponding to the  sub-sequence  $i_{1:t}=\{ i_1, i_2, \dots, i_{t} \}$;  taking an action  will lead to some \emph{reward} $r$. The key point is that we utilize the discriminator components to provide the reward signal for guiding the learning of the generator. We define the expected return $Q(s,a)$ for a pair of state and action, namely the $Q$-function, as below

\begin{equation}
\label{eq:action-value}
Q(s = i_{1:t-1}, a = i_t) = \sum_{j=1}^{m} \omega_j \hat{y}_j, 
\end{equation}
where $\hat{y}_j$ is the rationality score~(Eq.~\eqref{eq-mlp}) of current sequence according to the $j$-th discriminator, and $\omega_j$ is the combination coefficient defined through a $\lambda$-parameterized softmax function: 
\begin{equation}\label{eq-lambda}
\omega_j = \frac{\exp(\lambda\hat{y}_j)}{\sum_{j'=1}^m \exp(\lambda\hat{y}_{j'})},
\end{equation}
where $\lambda$ is a tuning parameter that will be discussed later.

As the discriminator is updated iteratively, it  gradually pushes the generator to its limit, which will generate more realistic recommended items.  Through multiple-factor enhanced architecture, the generator can obtain guidance of sequential characteristics in the interaction sequence from different perspectives.

\subsubsection{Learning Algorithm} After the task is formulated as a RL setting,  we can apply the classic policy gradient to learn the model parameters of the generator. Formally, the objective of the generator $G_\theta(i_t|i_{1:t-1})$ is to maximize the expected reward at the $t$-th step:
\begin{equation*}
\begin{split}
\mathcal{J}(\theta) = \mathbb{E}[R_t|i_{1:t-1}; \theta] = \sum_{i_t \in \mathcal{I}} G_\theta(i_t|i_{1:t-1}) \cdot Q(i_{1:t-1}, i_t),
\end{split}
\end{equation*} 
where $G_\theta(i_t|i_{1:t-1})$  and $Q(i_{1:t-1}, i_t)$ are defined in Eq.~\eqref{eq-predict} and Eq.~\eqref{eq:action-value}, respectively. 
$R_t$ denotes the reward of a generated sequence.
The gradient of the objective function $\mathcal{J}(\theta)$ \emph{w.r.t.} the generator's parameters $\theta$ can be derived as:
\begin{equation}
\label{eq:gradient}
\begin{split}
\nabla_\theta \mathcal{J}(\theta) 
&= \nabla_\theta \sum_{i_t \in \mathcal{I}} G_\theta(i_t|i_{1:t-1}) \cdot Q(i_{1:t-1}, i_t)\\
&= \sum_{i_t \in \mathcal{I}} \nabla_\theta G_\theta(i_t|i_{1:t-1}) \cdot Q(i_{1:t-1}, i_t)\\
&= \sum_{i_t \in \mathcal{I}} G_\theta(i_t|i_{1:t-1}) \nabla_\theta \log G_\theta(i_t|i_{1:t-1}) \cdot Q(i_{1:t-1}, i_t) \\
&= \mathbb{E}_{i_t \sim G_\theta(i_t|i_{1:t-1})}[\nabla_\theta \log G_\theta(i_t|i_{1:t-1}) \cdot Q(i_{1:t-1}, i_t)].
\end{split}
\end{equation}

We update the  parameters of the generator using gradient ascent as follows:
\begin{equation}
\begin{split}
\label{E2}
\theta \leftarrow \theta + \gamma \nabla_\theta \mathcal{J}(\theta),
\end{split}
\end{equation}
where $\gamma$ is the step size of the parameter update.

\ignore{$R_t$ is the reward for a complete sequence generated by generator. As mentioned before, the reward is from the discriminators $D_\Phi$ and denoted by action-value function $Q_{D, G}(s, a)$ of the sequence, measuring the expected value signal based on the state $s$ and action $a$.
Indeed, we take the estimated probability of being a reasonable sequence from multiple discriminators as the reward:
\begin{equation}
\begin{split}
\label{eq:action-value}
Q_{D,G}(a = I_{1:T-1}, s = i_T) 
&= softmax(D_{\Phi} (I_{1:T}), \lambda) \\
&= \sum_{j=1}^{|\Phi|} \omega_j D_{\phi_j} (I_{1:T}),
\end{split}
\end{equation}
where $\omega_j = {\exp{(\lambda D_{\phi_j} (I_{1:T}))}} / {\sum_{k} \exp{(\lambda D_{\phi_k} (I_{1:T})})}$ is the weight for the j-$th$ discriminator and $|\Phi|$ is the number of discriminators.
To simplify the notation, we use $\alpha_j$ to indicate $D_{\phi_j} (I_{1:T})$ and $\alpha$ to indicate $Q_{D,G}(s = I_{1:T-1}, a = i_T)$ in Fig.~\ref{fig-model}.
}

After updating the generator, we continue to optimize  each discriminator $D_{\phi_j}$ by minimizing the following objective loss:
\begin{equation}
\label{eq:opt_d}
\begin{split}
\min \limits_{\phi_j} - \mathbb{E}_{i_{1:t} \sim P_{\text{data}}}[\log D_{\phi_j} (i_{1:t})] - \mathbb{E}_{i_{1:t} \sim G_\theta}[\log (1 - D_{\phi_j} (i_{1:t})]\},
\end{split}
\end{equation}
where $P_\text{data}$ is the real data distribution.

Algorithm~\ref{alg1} presents the details of the training algorithm for our approach.
The parameters of $G_{\theta}$ and multiple discriminators $D_{\Phi}$ are pretrained correspondingly. 
For each $G$-step, we generate the recommended item based on  the previous sequence $i_{1:t-1}$, and then update the parameter by policy gradient with the reward provided from multiple discriminators.
For each $D$-step, the recommended sequence is considered as the negative samples and we take the actual sequence from training data as positive ones.
Then the discriminators are updated to discriminate between positive and negative sequences accordingly.
Such a process is repeated until the algorithm converges.

\begin{algorithm}[t] 
\caption{The learning algorithm  for our MFGAN framework.} 
\label{alg1} 
\begin{algorithmic}[1] 
\REQUIRE generator $G_\theta$; discriminators $D_\Phi = \{D_{\phi_1} ,  \dots, D_{\phi_{m}}\}$; user-item interactive sequence dataset $\mathcal{S}$ 
\STATE Initialize $G_\theta$, $D_\Phi$ with random weights $\theta$, $\Phi$
\STATE Pre-train $G_\theta$ using MLE
\STATE Generate negative samples using $G_\theta$ for training $D_\Phi$
\STATE Pre-train $D_\Phi$ via minimizing cross-entropy
\REPEAT 
    \FOR{$G$-steps} 
        \STATE Generate the predicted item $i_t$ using $i_{1:t-1}$
        \STATE Obtain the generated sequence $i_{1:t}$
        \STATE Compute $Q(s = i_{1:t-1}, a = i_t)$ by Eq.~ \eqref{eq:action-value}
        \STATE Update generator parameters $\theta$ via policy gradient Eq.~ \eqref{E2}
    \ENDFOR 
    
    \FOR{$D$-steps} 
    \STATE Use $G_\theta$ to generate negative examples
    \STATE Train $m$ discriminators $D_\Phi$ by Eq.~\eqref{eq:opt_d}
    \ENDFOR 
\UNTIL{Convergence} 

\end{algorithmic} 
\end{algorithm}

\subsection{Discussion and Analysis}\label{sec:theory}


In this section, we analyze the effectiveness and the stability of the multi-adversarial architecture in our task.

As mentioned before, we train the MFGAN model in an RL way by policy gradient. Since we have multiple discriminators, from each discriminator we receive a reward to guide the training process of the generator. 
Recall that we use a $\lambda$-parameterized softmax function to combine the reward signals from multiple discriminators in Eq.~\eqref{eq-lambda}. By using such a parameterization, our reward function can be implemented in several  forms: 

(1)  $\lambda\rightarrow -\infty$: it selects the discriminator with the minimum reward, \ie \emph{min};

(2)  $\lambda\rightarrow +\infty$: it selects the discriminator with the maximum reward, \ie \emph{max};

(3) $\lambda=0$: it becomes a simple average over all the discriminators, \ie \emph{mean}; 

(4) Others: it adopts a ``soft'' combination of multiple reward values. 

Among the four cases, we first study two extreme strategies, namely \emph{max} and \emph{min}.
As shown in \cite{ICLR17GMAN}, it is too harsh for the generator by adopting the minimum reward from the  discriminators.
  Let $p_G (x)$ denote the distribution induced by the generator. The low reward only indicates the position where to decrease $p_G (x)$, and does not specifically indicate the position where to increase $p_G (x)$. In addition, decreasing $p_G (x)$ will increase $p_G (x)$ in other regions of distribution space $\mathcal{X}$ ( keeping $\int_\mathcal{X} p_G (x) = 1$), and the correctness of this region cannot be guaranteed. Hence, the \emph{min} strategy is not good to train our approach. Conversely, \emph{max} always selects the maximum reward that is able to alleviate the above training issue of \emph{min}. However, since multiple factors are involved in the discriminators, some ``prominent'' factor will dominate the learning process, leading to insufficient training of other factors.

Compared with the former two cases, the latter two cases seem to be more reasonable in practice. 
They consider  the contributions from all the discriminators. 
Specially, we can have an interesting observation: the gradient $\nabla_\theta \mathcal{J}(\theta)$ of the generator calculated in Eq.~\eqref{eq:gradient} is more robust for model learning  due to the use of multiple discriminators. The major reason is that the $Q$-function in Eq.~\eqref{eq:action-value} is 
equal to zero \emph{if and only if}  $D_{\phi_j} = 0$ for all $j$, when all the discriminators give zero  reward.
Therefore, using multiple discriminators is indeed useful to stabilize the learning process of our approach. 
In our experiments, we do not observe a significant difference between the last cases on our task. 
Hence, we adopt the simpler \emph{mean} way to set our reward function.

Our work is closely related to general sequence prediction studies with adversarial training~\cite{Yu2017SeqGAN} or reinforcement learning~\cite{Ranzato2015Sequence}.  Similar to SeqGAN~\cite{Yu2017SeqGAN}, we set up two  components with different roles of \emph{generator} and \emph{discriminator}, respectively. 
Also, it is  easy to make an analogy between the two roles and the concepts of ``\emph{actor}'' and ``\emph{critic}'' in the actor-critic algorithm in RL~\cite{Ranzato2015Sequence}.
Compared with previous sequential recommendation models~\cite{GRU-f,HuangWSDM19, li2019review}, 
our approach takes a novel perspective that decouples various factors from the prediction component. 
In our model, each discriminator has been fed with the information of some specific factor.
Such a way is able to enhance the interpretability of the generator, that is to say,  the reward values of discriminators can be treated as the importance of influencing factors at each time step.   
With the proposed framework, we believe there is much room to consider more advanced implementations or functions for generator and discriminators for improving sequential recommendation.




\ignore{Note that the gradient $\nabla_\theta J(\theta)$ of the generator calculating in Eq.~\eqref{eq:gradient} is minimized at $\sum_{j=1}^{|\Phi|} D_{\phi_j} = 0$, if and only if $D_{\phi_j} = 0$ for all $j$. So generator only receives a vanishing gradient when all $D_{\phi_j}$ consider that the sample is unreasonable. For our multiple discriminators from different perspectives, the probability of occurrence is much smaller than the single discriminator without external factors. In other wards, the generator only needs to fool a single $D_{\phi_j}$ to receive a constructive feedback. This result allows the generator to successfully minimize the generator objective and generally guarantee the stability of the training process. We also verified the above conclusions in further experiments.
}
\section{Experiments}
\label{sec:exp}
In this section, we first setup the experiments,  then report major comparison results and other detailed analysis. 

\subsection{Dataset Construction}
We construct experiments on three public datasets from different domains, including \textsc{MovieLens-1M} movie~\cite{Movielens}, \textsc{Yahoo!} music\footnote{\url{https://webscope.sandbox.yahoo.com/catalog.php?datatype=r}.} and \textsc{Steam}~\cite{kang2018self} game. 
Since the \textsc{Yahoo!} dataset is very large, we randomly sample a subset of ten thousand users from the entire dataset. 
We group the interaction records by users, sort them by the timestamps ascendingly, and form the interaction sequence for each user.
Following \cite{RendleFS10}, we only keep the $k$-core dataset, filtering out unpopular items and inactive users with interaction records which are fewer than $k$. We set $k = 5, 10, 5$ for the \textsc{MovieLens-1M}, \textsc{Yahoo!} and \textsc{Steam} datasets, respectively. 

The three datasets contain several kinds of context information. We extract such context information as \emph{factors}:

 (1) For \textsc{MovieLens-1M} dataset, we select category, popularity, and knowledge graph information as factors. Note that we treat the knowledge base~(KB) information as a single factor, since we would like to develop a strong factor. 
We use the KB4Rec dataset~\cite{KB4Rec} to obtain \emph{item-to-entity} alignment mapping, and then obtain the KB information from \textsc{Freebase}~\cite{freebase}. We adopt the classic TransE~\cite{transe-NIPS-2013} to learn the factor representation for KB information. 

(2) For \textsc{Yahoo!} dataset, we select category, popularity, artist and album as factors. 

(3) For \textsc{Steam} dataset, we select category, popularity, price and developer as factors. 
\ignore{For the numeric factors like price, we divide all prices into multiple ranges and use it to indicate the price level.}

Note that for all the three datasets, we have incorporated a special kind of factor, \ie item ID. Although it has been utilized in the generator, it can be utilized by the discriminator in a different way, \ie using a bi-directional Transformer architecture. 
Item ID can be used as a reference to compare the usefulness of the other kinds of factor information. 
Following~\cite{kang2018self}, we split each user interaction sequence into three parts:  the last item in the interaction sequence is treated as test data, the item just before the last is used for validation and the rest data is considered  as  training data.
The statistics of three datasets after preprocessing are summarized in Table~\ref{tab-data}.

\ignore{Each dataset has its own factors, in movielens-1m datasets, we leverage category, popularity, and knowledge graph information, in music dataset we leverage category, popularity, artist and album, and we utilize category, popularity, price and developer in the game dataset. We also treat origin identifier of item as a factor which is not external context information.
As we adopt knowledge graph information as a factor in movielens-1m dataset, we link the filtered items with \textsc{Freebase}~\cite{freebase} entities using KB4Rec dataset~\cite{KB4Rec}, which contains the alignment records between Freebase entities and items from several domains. We only keep the interactions related to the linked items in the final dataset and we apply TransE~\cite{transe-NIPS-2013} to obtain the entity embeddings from knowledge graph. To train TransE algorithm, we start with linked entities as seeds and expand the graph with one-step search.}

\begin{table}[htbp]
  \centering
  \caption{Statistics of the three datasets.}
  \label{tab-data}%
    \begin{tabular}{c r r r r}
		\hline
		Dataset & \#Users & \#Items & \#Interactions & \#Factors\\
		\hline
		\hline
		Movielens-1m   & 6,040 	& 3,361     & 996,834   & 4\\
		Music   & 10,000     & 136,630   & 3,732,463 & 5\\
		Game   & 51,995 	& 12,037    & 1,789,953 & 5\\
		\hline

    \end{tabular}%
\end{table}%

\begin{table*}[h]
	\centering
	\caption{Performance comparison of different methods for sequential recommendation task on three datasets. We use bold and underline fonts to denote the best performance and second best performance method in each metric respectively.}
	\label{tab:results-all}%
	\begin{tabular}{| p{1.3cm} | p{1.3cm}<{\centering} p{1.3cm}<{\centering} p{1.3cm}<{\centering}| p{1.3cm}<{\centering} p{1.3cm}<{\centering} p{1.3cm}<{\centering}| p{1.3cm}<{\centering} p{1.3cm}<{\centering} p{1.3cm}<{\centering}|}
		\hline
		\multirow{2}{*}{Models}&\multicolumn{3}{c|}{Movie}&\multicolumn{3}{c|}{Music}&\multicolumn{3}{c|}{Game}\\
		\cline{2-10}
		&NDCG@10    & HR@10 & MRR&NDCG@10    & HR@10 & MRR&NDCG@10    & HR@10 & MRR\\
		\hline
		\hline
		PopRec  & 0.2290 & 0.4200 & 0.1940 & 0.3984 & 0.6322 & 0.3408 & 0.4129 & 0.6738 & 0.3477\\
		BPR     & 0.3063 & 0.5576 & 0.2505 & 0.4538 & 0.6695 & 0.3992 & 0.4203 & 0.6919 & 0.3511\\
		FM      & 0.1935 & 0.3522 & 0.1690 & 0.2196 & 0.4165 & 0.1812 & 0.1383 & 0.2818 & 0.1216\\
		IRGAN   & 0.2263 & 0.4201 & 0.1903 & 0.2857 & 0.5120 & 0.2332 & 0.3942 & 0.6525 & 0.3307 \\
		\hline
		FPMC 	& 0.3754 & 0.5990 & 0.3237 & 0.4330 & 0.5768 & 0.4022 & 0.3699 & 0.6167 & 0.3107\\
		GRU 	& 0.4014 & 0.6233 & 0.3763 & 0.4843 & 0.7778 & 0.4053 & 0.2383 & 0.4212 & 0.2037\\
		GRU$_F$ & 0.4130 & 0.6409 & 0.3552 & 0.5470 & 0.8112 & 0.3961 & 0.2806 & 0.4860 & 0.2369 \\
		SASRec  & \underline{0.5849} & \underline{0.8132} & \underline{0.5214} & \underline{0.8356} & \underline{0.9293} & \underline{0.8074} & \underline{0.5895} & \underline{0.8351} & \underline{0.5203} \\
		\hline
		MFGAN & \textbf{0.6185} & \textbf{0.8318} & \textbf{0.5587} & \textbf{0.8556} & \textbf{0.9417} & \textbf{0.8286} & \textbf{0.6035} & \textbf{0.8488} & \textbf{0.5346} \\
		\hline
	\end{tabular}%
\end{table*}%

\subsection{Experimental Settings}

\subsubsection{Evaluation Protocol}
To assess whether our method can improve the sequential recommendation, we adopt a variety of evaluation metrics widely used in previous works~\cite{RendleFS10, Huang-2018-SIGIR, HuangWSDM19}: Mean Reciprocal Rank (MRR), top-$k$ Normalized Discounted cumulative gain (NDCG@10) and Hit Ratio (HR@10). 
Since the item set is large, it is time-consuming to enumerate all the items as candidate. For each positive item in the test set, we pair it with 100 sampled items that the user has not interacted with as negative items. The evaluation metrics can be computed according to the rankings of these items. 
We report the average score over all test users.

\subsubsection{Implementation Details}
For our proposed MFGAN, we use two self-attention blocks in the generator and one self-attention block in discriminators. In both generator and discriminators, item embeddings are shared in embedding layer and prediction layer. We implement MFGAN with \emph{Tensorflow}, and use \emph{Adam} optimizer. At the pre-training stage, we first train the generator and discriminators to converge separately. At the adversarial training stage, we alternatively train the generator and discriminators with 100 epochs and 1 epoch, respectively. The learning rates   are set to 0.001 for \textsc{MovieLens-1M} dataset, and 0.0002 for \textsc{Steam} and \textsc{Yahoo!} datasets, and the dropout rates are all set to 0.2 for the three datasets.
For each factor, we discretize its possible values into several bins if needed. Each bin is associated with an embedding vector. 
The item embedding size and all the factor embedding sizes are set to 50, the KB embedding size with \textsc{TransE} is set to 100.
The mini-batch sizes are set to 128 in generator and 16 in discriminators.


\subsubsection{Comparison Methods}

Here, we compare our propose approach MFGAN against a number of competitive baselines. 
Our baselines include related methods on general and sequential recommendation with or without context information:

$\bullet$ \textbf{PopRec}: This is a method that sorts items by their popularity to assign each item a rank, tending to recommend highly popular items to users.

 $\bullet$ \textbf{BPR}~\cite{Rendle-09-bpr}: This is a classic personalized ranking algorithm that optimizes the pairwise ranking loss function of latent factor model with implicit feedback.

$\bullet$ \textbf{FM}~\cite{FM2012}: It uses a generic matrix factorization to learn the coefficients of combined features, considering the interactions between different features.

$\bullet$ \textbf{IRGAN}~\cite{IRGAN}: This method combines generative and discriminative information retrieval via adversarial training, in which a simple matrix factorization is used for the discriminator.

$\bullet$ \textbf{FPMC}~\cite{RendleFS10}: It combines Matrix Factorization and Markov Chain, which can simultaneously capture sequential information and long-term user preference.

$\bullet$ \textbf{GRU}~\cite{GRU4Rec}: It is a GRU-based sequential recommender with session-parallel mini-batch training, which employs ranking-based loss functions. We implement an enhanced version by replacing one-hot vectors with pre-trained BPR vectors.

$\bullet$ \textbf{GRU$_F$}~\cite{GRU-f}: It proposes to incorporate additional feature vector as the input of GRU networks, which incorporates auxiliary features to improve sequential recommendation.

$\bullet$ \textbf{SASRec}~\cite{kang2018self}: It is a next-item sequential recommendation method based on the Transformer architecture, which adaptively considers interacted items for prediction. This method is the state-of-the-art baseline for sequential recommendation. 

$\bullet$ \textbf{MFGAN}: This is our approach introduced in Section~\ref{sec:method}.

Besides GRU$_F$,  there are other baselines that utilize context information. However,  we have empirically found that the Transformer architecture is more superior than other sequence neural networks in sequential recommendation. Indeed, SASRec is significantly better than quite a few context-aware recommendation algorithms with other RNN architectures. 
Therefore, our focus is how our approach improves over SASRec, and we omit other context-aware  baselines. Another note is that in comparison our MFGAN approach only utilizes the generator to recommend items, \ie context information or discriminators will not be used at the test stage. 

\subsection{Results and Analysis}

The results of different methods for sequential recommendation are presented in Table~\ref{tab:results-all}. It can be observed that:

(1) Among non-sequential recommendation baselines, we can find that BPR outperforms other methods, and  non-sequential recommendation methods overall perform worse than sequential recommendation methods.
Specially, factorization machine~(FM) does not work very well in this task, since it still adopts the regression loss and cannot effectively capture the preference order over two items by a user. Besides, popularity seems to be a robust baseline that gives substantial performance on our datasets. A major reason is that there may exist the ``\emph{rich-gets-richer}'' phenomenon in product adoption, and using $k$-core preprocessing on our datasets further enhances this trend.

(2) Among sequential recommendation baselines, the Markov Chain-based method performs better on sparse dataset (\ie the game dataset) than dense datasets (\ie the other two datasets).
As for two neural methods, GRU adopts the RNN-based architecture, serving as a standard comparison method for sequential recommendation. With available context information, GRU$_F$ further improves over the GRU method, indicating context information is useful in our task.
Furthermore, SASRec utilizes the powerful Transformer architecture to develop the sequential recommender, achieving the best performance among all the baselines. 
In natural language processing, self-attention architecture has shown its superiority in various tasks~\cite{bert, transformer}. Such an architecture is particularly useful when dealing with sequence data, which also works well in sequential recommendation. 

(3) Finally, we compare our proposed model MFGAN with  the baseline methods. It is clear to see that MFGAN is consistently better than these baselines by a large margin. Our base architecture is a pre-trained self-attentive generator. Different from the above models, we adopt the multi-adversarial architecture to guide the learning of the generator with various kinds of factor information. Note that 
our generator has not directly used any context information, and it is guided by the signals from the  discriminators. 
 Such a comparison indicates the multi-adversarial training approach is effective to improve the performance of self-attention architecture for sequential recommendation.

\begin{figure*}[ht]
	\centering
	\subfigure[\textsc{Movielens-1m} movie dataset.]{\label{fig-ablation-movie}
		\centering
		\includegraphics[width=0.25\textwidth]{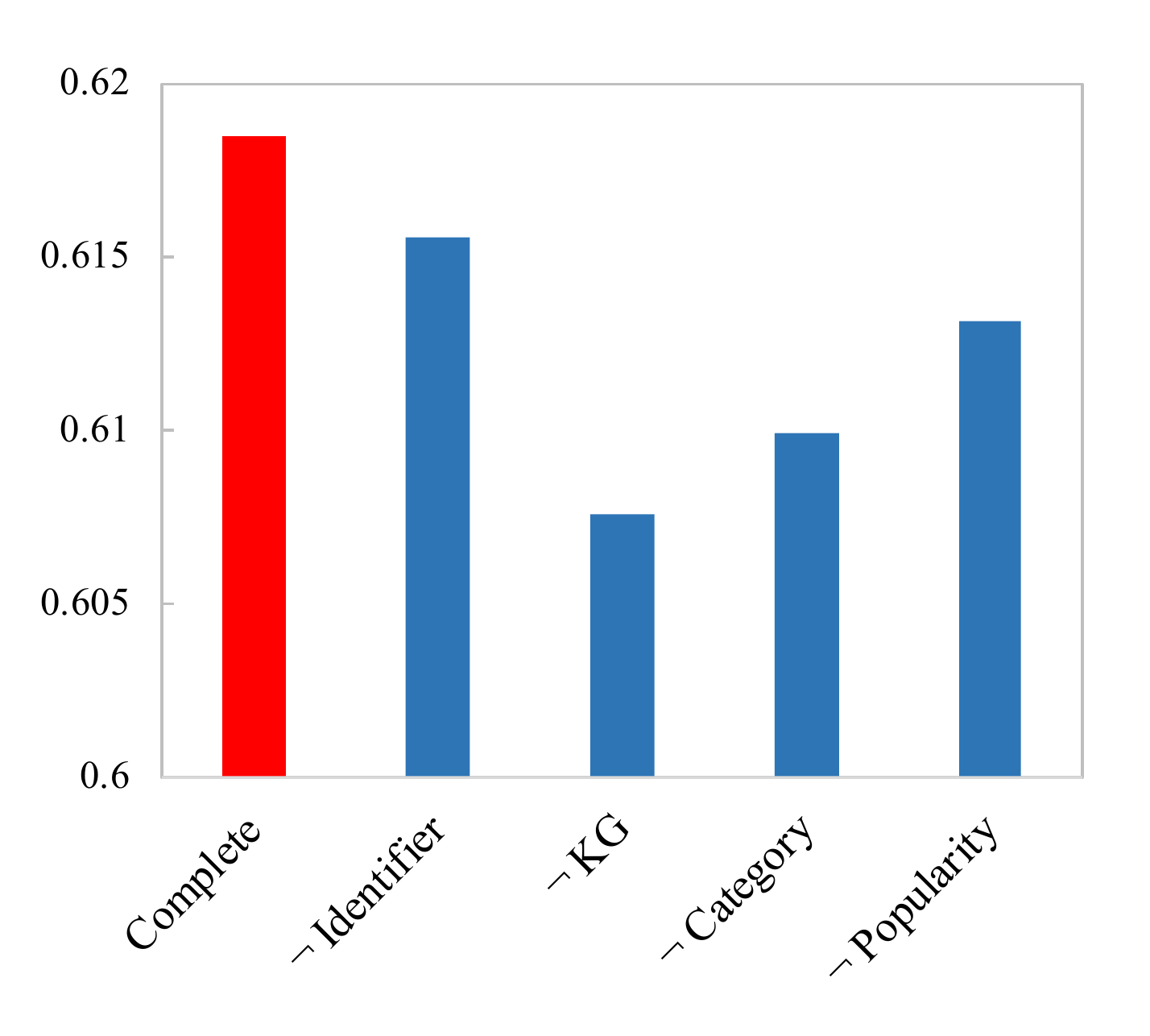}
	}
	\subfigure[\textsc{Yahoo!} music dataset.]{\label{fig-ablation-music}
		\centering
		\includegraphics[width=0.25\textwidth]{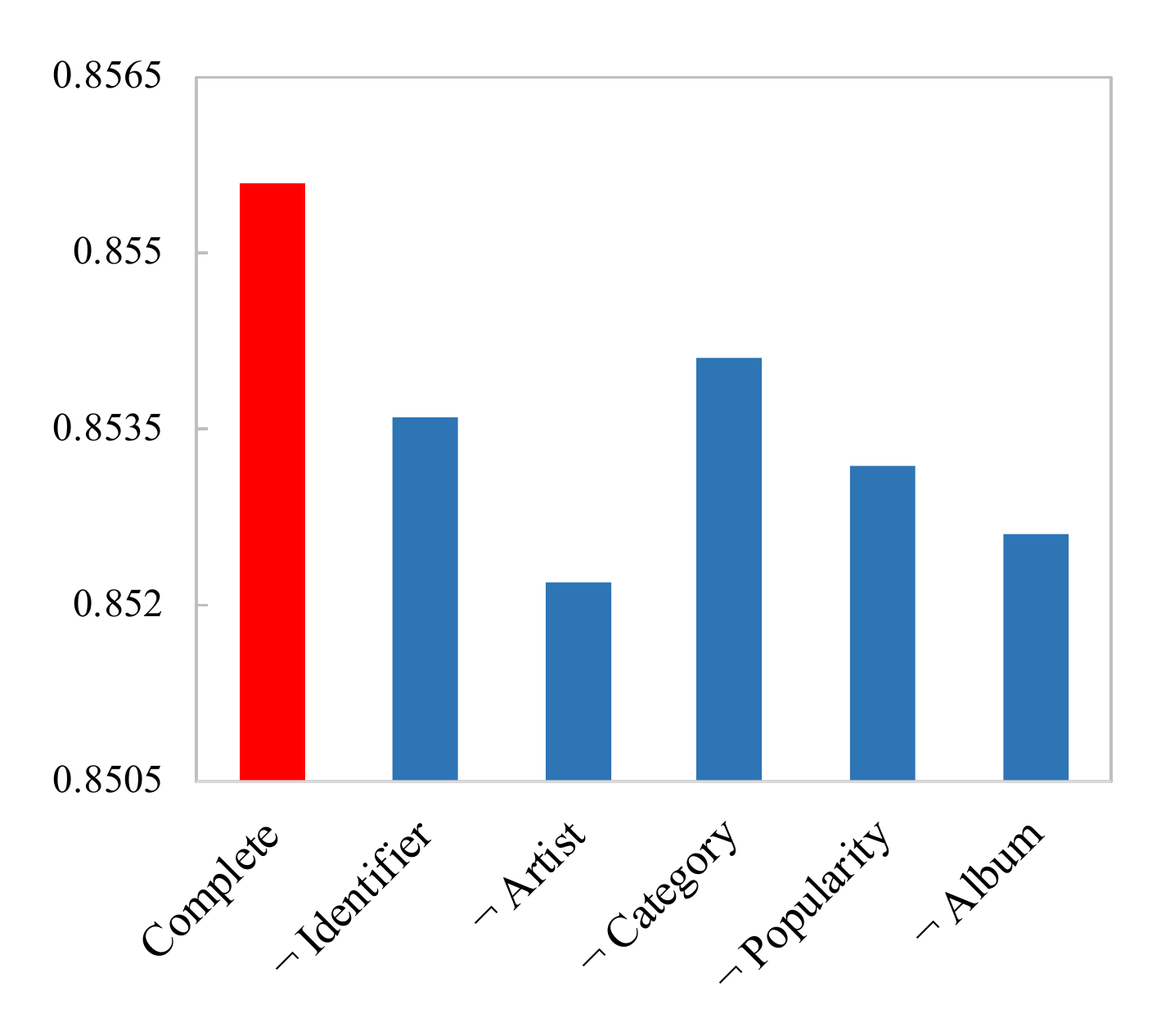}
	}
	\subfigure[\textsc{Steam} game dataset.]{\label{fig-ablation-game}
		\centering
		\includegraphics[width=0.25\textwidth]{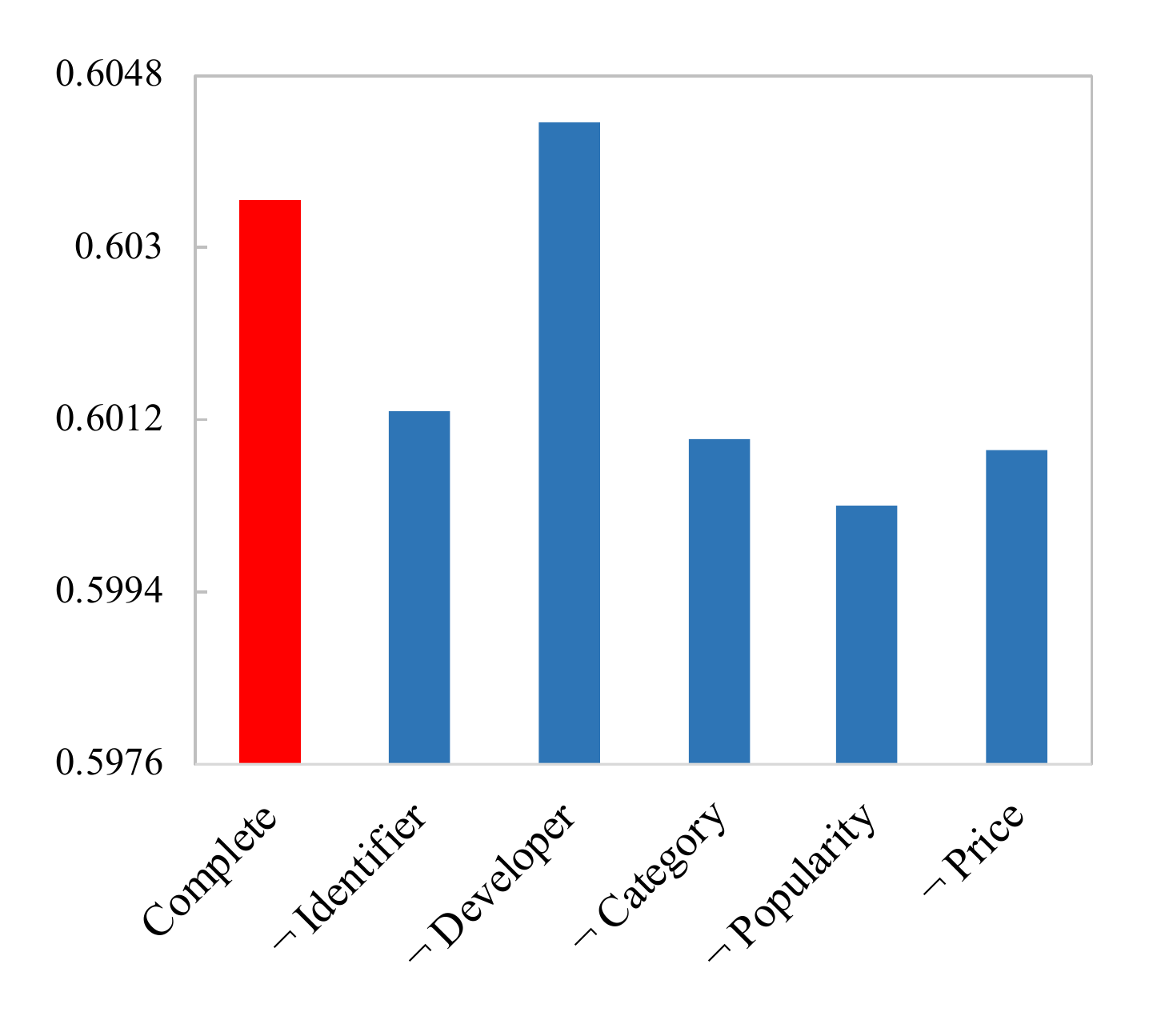}
	}
	\caption{Effect of different kinds of factor information in our framework for sequential recommendation. We report the results using the NDCG@10 metric. ``$\neg$'' indicates that we remove the corresponding factor information, while the rest factor information is kept. }
	\label{fig-ablation}
\end{figure*}

\subsection{Detailed Analysis on Our Approach}

In this section, we further conduct a series of detailed experiments to analyze the effectiveness of our approach.  

\subsubsection{Effect of Different Factors}
In this part, we study the effect of different kinds of factor information in our approach. 
We first prepare the complete MFGAN model, and then remove one kind of factor information at each time. 
In this way, we can study how each individual factor contributes to the final performance for sequential recommendation. 
Figure~\ref{fig-ablation} presents the NDCG@10 results of factor removal experiments for MFGAN. 
As we can see, all the factors seem to be useful to improve the performance of our MFGAN, except the factor \emph{developer} on the \textsc{Steam} game dataset. The most useful factors are \emph{knowledge graph}, \emph{artist} and \emph{popularity} for the three datasets, respectively. 
Interestingly,  item ID is not always  the most useful feature for discriminators, indicating the importance of other kinds of factor information.

\subsubsection{Comparisons of Model Variants}
In our model, we propose several techniques to improve the performance of our approach. Here, we construct  comparison experiments to examine the effect of these techniques:
 (1) \underline{\emph{Uni-directional}} using a uni-directional self-attention architecture in discriminators; 
 (2) \underline{\emph{SDSF}} using  a single discriminator considering  item ID as the only factor; (3) \underline{\emph{SDAF}} using a single discriminator incorporating all factor embeddings using simple vector concatenation.
The first variant is used to illustrate the effect of the bi-directional architecture, 
and the last two variants are used to illustrate the effect of using multiple discriminators. 

\begin{table}[htbp]
  \centering
  \caption{Variant comparisons of our MFGAN framework on \textsc{Movielens-1m} dataset.}
  \label{tab:ablation}%
    \begin{tabular}{l p{1.3cm}<{\centering} p{1.3cm}<{\centering} p{1.3cm}<{\centering}}
  \toprule
  Architecture            & NDCG@10 & HR@10 & MRR \\
  \midrule
  MFGAN                 & \textbf{0.6185} & \textbf{0.8318} & \textbf{0.5587} \\
  \hline
  SASRec                & 0.5849 & 0.8132 & 0.5214 \\
  Uni-Directional       & 0.6158 & 0.8299 & 0.5558 \\
  SDSF                  & 0.6032 & 0.8212 & 0.5413 \\
  SDAF                  & 0.6039 & 0.8221 & 0.5455 \\
  \bottomrule
    \end{tabular}%
\end{table}%

\begin{figure}[!tb]
	\small
	\centering 
	\includegraphics[width=0.37\textwidth]{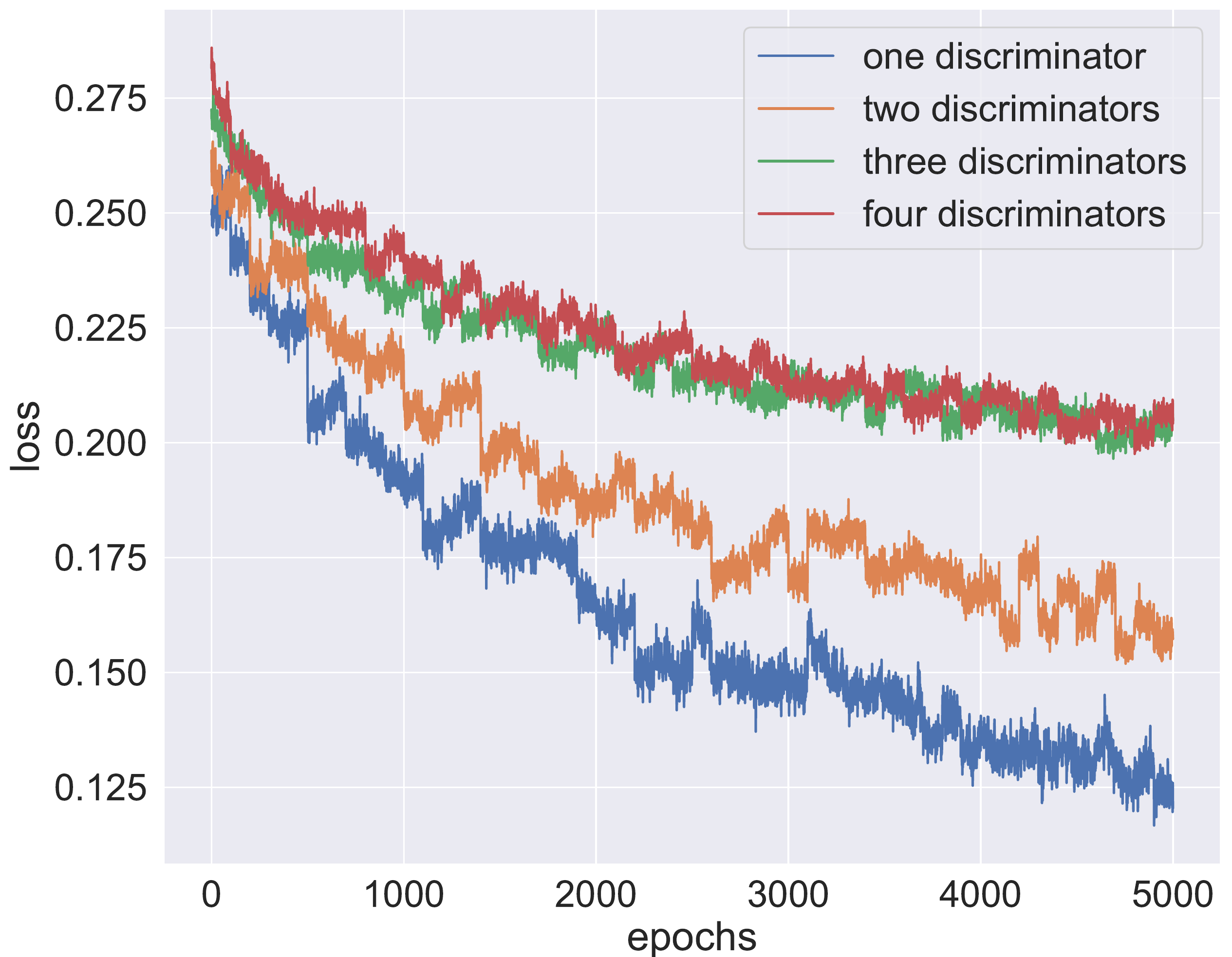}
	\caption{Convergence of loss with different numbers of discriminators in \textsc{Movielens-1m} dataset.} 
	\label{fig-loss} 
\end{figure}

Table~\ref{tab:ablation} presents the comparisons of these variants against our complete approach and SASRec. We can see that all the variants are worse than the complete approach. It indicates that  the bi-directional architecture and  multi-discriminator adversarial training are effective to improve the performance.
However, the improvement with bi-directional architecture seems to be small in numerical values.
A possible reason is that the uni-directional Transformer architecture  is ready very competitive. 
While,   the idea that utilizes the bi-directional architecture in discriminators has important implications. As future work, it will be interesting  to design other architectures for generator and discriminator, respectively. 

\begin{figure*}[ht]
	\centering
	\subfigure[A case from \textsc{Movielens-1m} movie dataset.]{\label{fig-case-movie}
		\centering
		\includegraphics[width=0.49\textwidth]{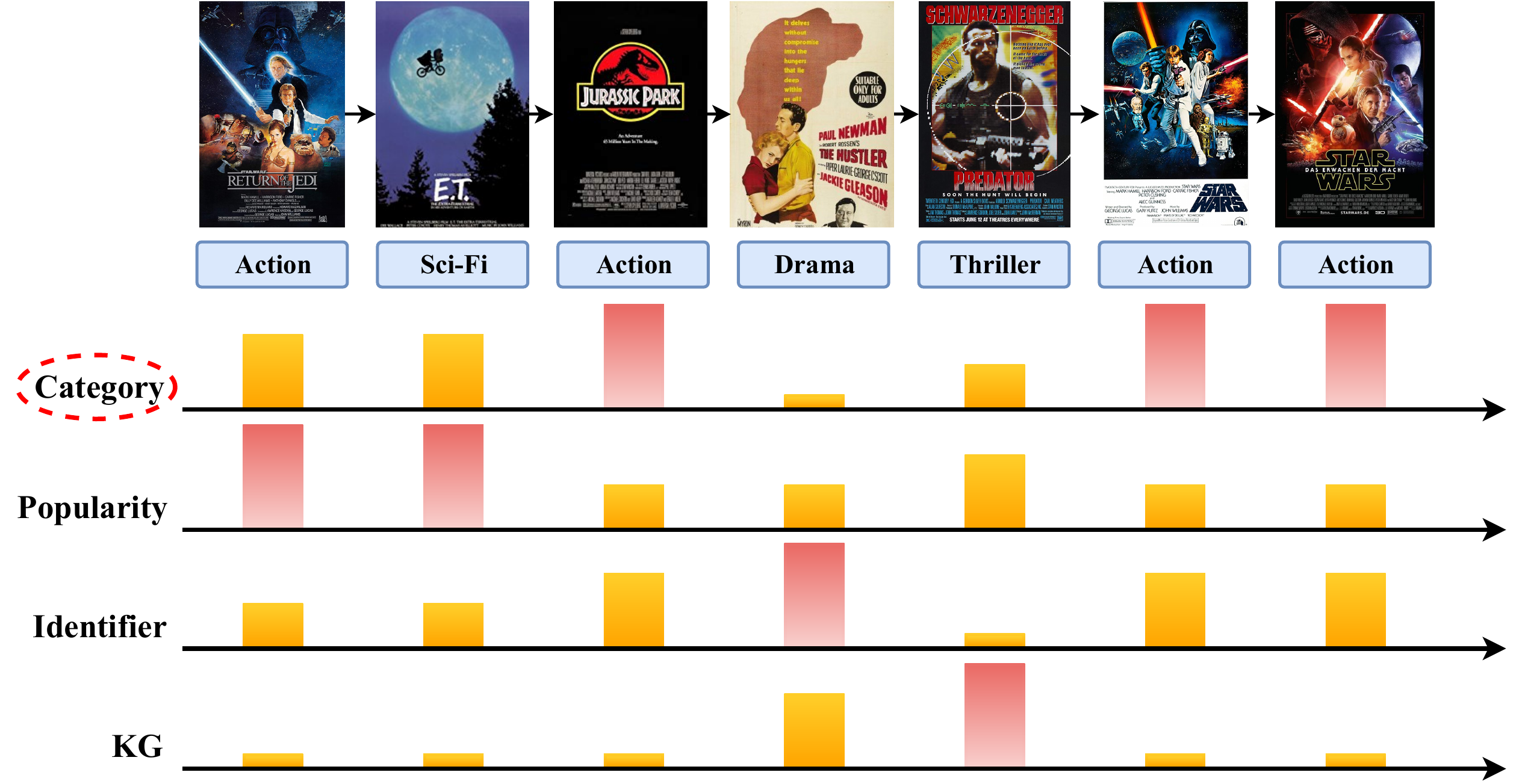}
	}
	\subfigure[A case from \textsc{Yahoo!} music dataset.]{\label{fig-case-music}
		\centering
		\includegraphics[width=0.48\textwidth]{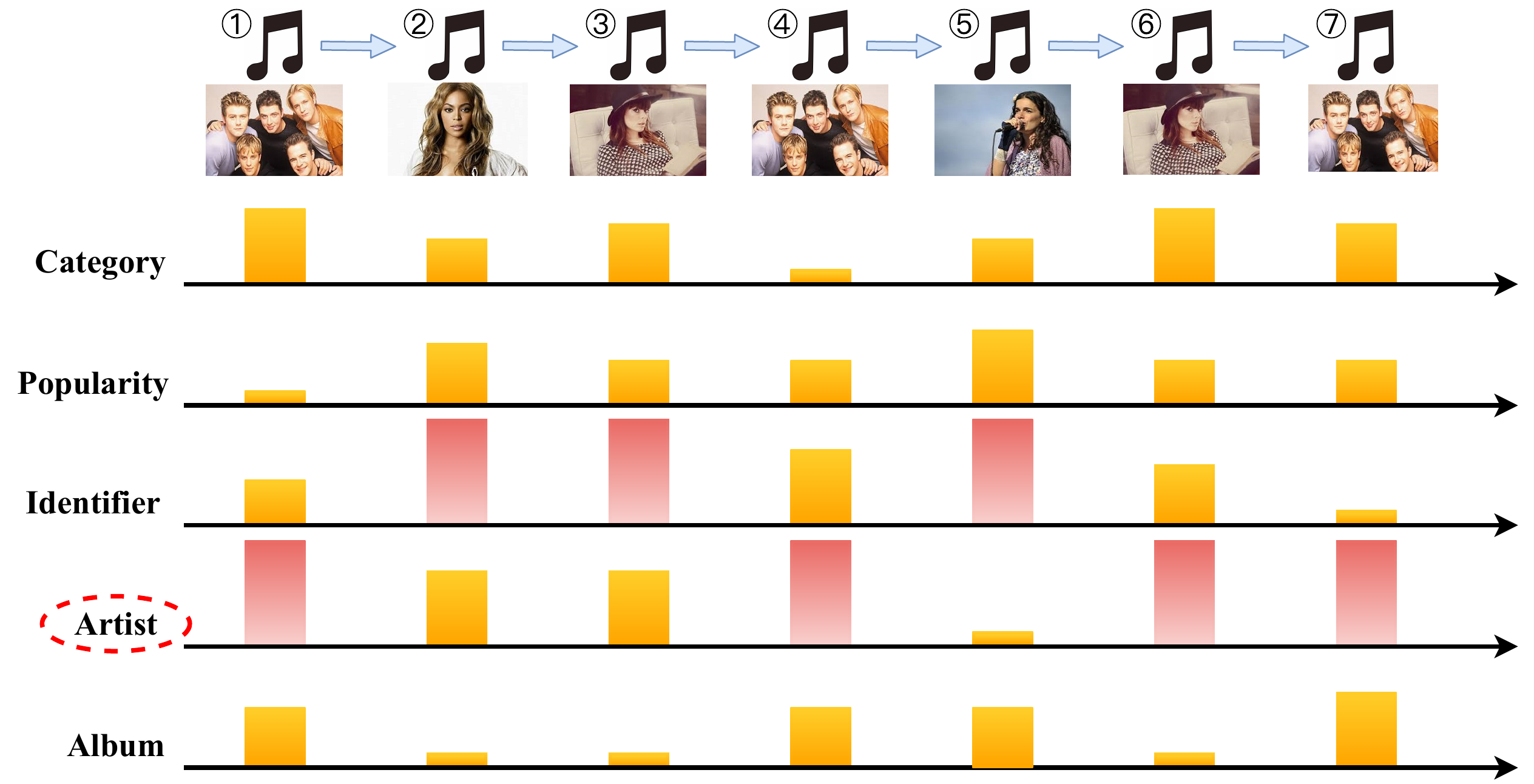}
	}
	\caption{Two qualitative cases with our approach. In each figure, the upper part is the interaction sequence, and the lower part corresponds to the lines of reward signal scores from some factors. 
	The height of the histogram represents the reward score within the interval $(0,1)$. We use the red color to highlight the highest bar (\ie the dominant factor) at each time step. For the two cases, we mainly focus on the \emph{category} and \emph{artist} factor, respectively. }
	\label{fig-case}
\end{figure*}

Note that the \emph{SDSF} variant can be considered as an enhanced implementation of SeqGAN~\cite{Yu2017SeqGAN}, where the generator and the discriminator have been  improved using the Transformer architecture. SeqGAN is originally proposed for general sequence generation, while it is easy to adapt to our task. 

Besides the performance improvement, as discussed in Section~\ref{sec:method}, using multiple discriminators is likely to stabilize the training of adversarial learning approaches.  
For this purpose, we further compare the convergence of our objective function with a varying number of discriminators during training process.
As we can see in Fig.~\ref{fig-loss}, using more discriminators is faster to achieve a relatively stable loss. The red and green lines (corresponding to use three or four discriminators) become stable after about 2000 epochs, while the other lines are not stable even after 4000 epochs. Another interesting observation is that using more discriminators seems to yield a larger loss. We speculate that it has the similar effect of regularization, which prevents the model parameters from overfitting on training data. 

\subsection{Qualitative Analysis on the Recommendation Interpretability}
Previous experiments have verified that our model is able to generate high-quality sequential recommendations. Another major benefit of our work is that our recommendations are highly interpretable due to the decoupled factors in the multi-adversarial framework. 
Here, we present two illustrative examples in Fig.~\ref{fig-case}, showing how our approach improves the recommendation interpretability with decoupled factors.

The first case (Fig.~\ref{fig-case-movie}) presents an interaction sequence of a sample user from \textsc{Movielens-1m} dataset. As we can see, the first two movies ``\emph{Star War: episode VI}'' and ``\emph{E.T. the Extra-Terrestrial}'' have received  higher reward signals from \emph{popularity}, which is the prominent factor at the beginning stage. Interestingly, seen from the \emph{category} line,  the $3$-rd, $6$-th and $7$-th movies are mainly driven by the category factor. Given the the last two movies ``\emph{Star War: episode IV}'' and ``\emph{Star War: episode V}'', the user, influenced by early popularity, seems to be a Star Wars fan or an action movie fan.

\ignore{Fig.~\ref{fig-case-movie} presents an interaction sequence of a sample user from movielens-1m. Firstly, the user started with the movie \emph{Star War: episode VI}, which is an action movie. Then, he watched a Sci-Fi movie. The popularity of the first two movies is relatively high, so the model believes that the selection of these two movies is dominated by popularity.
For the third movie, he switched to \emph{Jurassic Park}, which is also an action movie, at this timestamp, the model captures through the feedback of the multiple discriminators that the user chose \emph{Jurassic Park} because of the category of the movie. 
Subsequently, the fourth and fifth movies are drama and thriller. For the last two movies, the user selected \emph{Star War: episode IV} and \emph{Star War: episode V} in turn, our model finds out exactly that he most likely chose these two movies because of the category, and the interesting point is that we can find this user is likely to be a Star Wars fan or an action movie fan.}

The second case (Fig.~\ref{fig-case-music}) presents an interaction sequence of a sample user from \textsc{Yahoo!} music dataset. 
Similarly, we can check the effect of multiple factors at different time steps. 
Here, we mainly analyze the \emph{artist} factor,  focusing on the 4-th, 6-th and 7-th songs.
Along the interaction sequence, this user switches among different artists. Our model is able to capture such listening patterns: 
when the user listens to a song of a previous artist (\ie previously appearing in the interaction sequence) occurs, the corresponding \emph{artist} factor becomes dominant in our model. 

As a comparison, existing methods mainly focus on integrating various kinds of context information into  unified representations. The two examples have shown the advantage of decoupling various factors for sequential recommendation. 

\ignore{The user started with three songs by different artists, note that the first artist is one of the user's frequent artists, our model thinks that the user listened to the song because of the first artist. The second artist and the third artist are not the artists he often listens to. This user is accustomed to listening to a few artists and rarely to others, when he listens to the songs by a frequent artist, the model generally thinks that the artist factor is dominant when he listens to the current song. When he does not listen to a fixed artist's song, it will considered to be dominated by identifier co-occurrence of the song.
Once the user listens to a song by an infrequent artist, the model will capture that artist is the dominant factor sensitively.

Through the illustrations above, we demonstrate that by decoupling the interactive information through multiple discriminators, we can know which factor is dominated of the user when selecting the item at each timestamp, and we can also know which factors are considered more when recommending items to a user.
}



\section{Related Work}
\label{sec:related}
 In this section, we review studies closely related to our work in two aspects.

\eat{
In this section, we review several lines of related works, including sequential models, and generative adversarial network.}

\eat{
\paratitle{General Recommendation.} 
Early recommendation systems were designed to model users and items with implicit feedback~\cite{KorenBV09} or explicit feedback~\cite{Rendle-09-bpr}. A classical method is collaborative filtering (CF), using the interests of a group with similar interests to recommend information that users are interested in~\cite{HerlockerKBR99, SarwarKKR01}. Moreover, FM~\cite{FM2012} can imitate most decomposition models through feature engineering.
Due to the success in some areas, deep learning techniques have been introduced to recommendation system. A typical work is NCF~\cite{NCF17}, which uses DNN to simulate noise implicit feedback signals. AutoRec~\cite{SedhainMSX15} uses the autoencoder model to predict missing ratings in the user-item matrix.
}

\paratitle{Sequential Recommendation.}
Early works on sequential recommendation are mainly based on Markov Chain (MC) assumption.
For instance, Rendle et al.~\cite{RendleFS10} fuse the matrix factorization and first-order Markov Chain for modeling global user preference and short-term interests, respectively.
Another line to model user behaviors is resorting to the recurrent neural network, which has achieved  great success on sequential modeling in a variety of applications.
Hidasi et al.~\cite{GRU4Rec} firstly introduce Gated Recurrent Units (GRU) to the session-based recommendation.
A surge of following variants modify it by introducing pair-wise loss function~\cite{GRU-f}, attention mechanism~\cite{LiRCRLM17}, memory network~\cite{chen2018sequential}, hierarchical structure~\cite{QuadranaKHC17}, etc.
Other architectures or networks have also been used~\cite{lv2019sdm,ma2019hierarchical}, achieving good performance.
Moreover, context information is often used to improve the recommendation performance and interpretability~\cite{GRU-f, wang2017context, Huang-2018-SIGIR, HuangWSDM19}.
Recently, self-attention network has achieved significant improvement in a bunch of NLP tasks~\cite{transformer, bert} and inspired a new direction on applying the self-attention mechanism to sequential recommendation problem~\cite{kang2018self}.

\ignore{
\paratitle{Explainable Recommendation.} 
Explainable recommendation refers to personalized recommendation algorithms that address the problem of why~\cite{DBLP:journals/corr/abs-1804-11192}. The term of explainable recommendation was formally introduced in recent years~\cite{ZhangL0ZLM14}, and a large number of explainable recommendation approaches have been proposed and applied in real-world systems, simply classified as matrix factorization explainable recommendation~\cite{ZhangL0ZLM14}, graph-based explainable recommendation~\cite{HeckelVPD17}, knowledge-based explainable recommendation~\cite{Huang-2018-SIGIR, WangWX00C19}, rule mining explainable recommendation~\cite{BalogRA19}, etc.
}

\paratitle{Generative Adversarial Network.} 
Generative Adversarial Network (GAN)~\cite{GAN-2014-NIPS} was originally proposed to mimic the generation process of given data samples.
Typically, GAN consists of two components: the generative model learns to map from a latent space to a data distribution of interest, while the discriminative model distinguishes candidates produced by the generator from the true data distribution.
A surge of follow-up works either improve the GAN framework by introducing more advanced training strategies, like f-divergence~\cite{f-gan}, Wasserstein distance~\cite{WuHTAG18}, MMD constraints~\cite{li2017mmd} or explore diverse applications under GAN framework~\cite{SRGAN,Yu2017SeqGAN,ZhangGFCHSC17}.
In recommender systems, the idea of GAN has already been explored to some extent. 
IRGAN~\cite{IRGAN} is firstly proposed to unify the generative model and discriminative model in the field of information retrieval.
Chae et al.~\cite{CFGAN} follow this line and further improve the training scheme by sampling a real-valued vector instead of a single item index.
Moreover, personalized determinantal point process is utilized to improve recommendation diversity via adversarial training~\cite{WuLMZZG19}.

Our work is related to the above studies, while has a different focus. 
We are the first to apply adversarial training to sequential recommender systems. 
By designing a multi-adversarial network, our model is able to explicitly characterize the effect of each individual factor on sequential recommendation over time, which improves the recommendation interpretability.

\section{Conclusions}
\label{sec:conclusions}

In this paper, we have proposed a Multi-Factor Generative Adversarial Network (MFGAN) for sequential recommendation. In our framework, the generator taking user behavior sequences as input is used to 
generate possible next items, and multiple factor-specific discriminators are used to evaluate the generated sub-sequence from the perspectives of different factors. We have constructed extensive experiments
on three real-world datasets. Experimental results have shown that our approach outperforms several competitive baselines. Especially, we have found that using multiple discriminators is useful to stabilize the training of adversarial learning, and also enhance the interpretability of recommendation algorithms.

Currently, we consider a simple setting where multiple discriminators are separately designed. As future work, we will investigate how to design a more principled way to share discrimination information across different discriminators. We will also consider incorporating explicit user preference in the discriminators.  

\section{acknowledgement}
This work was partially supported by the National Natural Science Foundation of China under Grant No. 61872369 and 61832017, the Fundamental Research Funds for the Central Universities, the Research Funds of Renmin University of China under Grant No.18XNL
G22 and 19XNQ047,   Beijing Academy of Artificial Intelligence (BAAI) under Grant No. BAAI2020ZJ0301, and Beijing Outstanding Young Scientist Program under Grant No. BJJWZYJH012019100020098. Xin Zhao is the corresponding author.

{
\bibliographystyle{ACM-Reference-Format}
\bibliography{with_ruiyang}
}

\end{document}